\documentclass[sigplan,screen]{acmart}

\setcopyright{acmcopyright}
\acmPrice{15.00}
\acmDOI{10.1145/3211346.3211352}
\acmYear{2018}
\copyrightyear{2018}
\acmISBN{978-1-4503-5834-7/18/06}
\acmConference[MAPL'18]{2nd ACM SIGPLAN International Workshop on Machine Learning and Programming Languages}{June 18, 2018}{Philadelphia, PA, USA}
\startPage{1}

\usepackage{color}
\usepackage{listings}
\usepackage{paralist}
\usepackage{amsmath}
\usepackage{mathtools}
\usepackage{textcomp}
\usepackage{xspace}
\usepackage{booktabs}
\usepackage{pbox}
\usepackage{subcaption}
\usepackage{paralist}
\usepackage{graphicx}
\usepackage{multirow}
\usepackage{amssymb}
\usepackage{epstopdf} 
\usepackage{verbatim}
\usepackage{comment}
\usepackage{framed}
\usepackage{enumitem}
\usepackage{flushend}

\newcommand{\Space}[1]{}

\let\norm\undefined 
\DeclarePairedDelimiter\norm{\lVert}{\rVert}

\definecolor{javared}{rgb}{0.6,0,0} 
\definecolor{javagreen}{rgb}{0.25,0.5,0.35} 
\definecolor{javapurple}{rgb}{0.5,0,0.35} 
\definecolor{javadocblue}{rgb}{0.25,0.35,0.75} 

\lstdefinestyle{customc}{
  belowcaptionskip=\baselineskip,
  breaklines=true,
  xleftmargin=\parindent,
  language=java,
  showstringspaces=false,
  basicstyle=\scriptsize\ttfamily,
  keywordstyle=\bfseries\color{javapurple},
  commentstyle=\itshape\blue,
  belowskip=-10pt,
  aboveskip=-5pt
}

\newcounter{RQCounter}
\newcounter{RQACounter}

\definecolor{gray50}{gray}{.5}
\definecolor{gray40}{gray}{.6}
\definecolor{gray30}{gray}{.7}
\definecolor{gray20}{gray}{.8}
\definecolor{gray10}{gray}{.9}
\definecolor{gray05}{gray}{.95}

\newlength\Linewidth
\def\findlength{\setlength\Linewidth\linewidth
\addtolength\Linewidth{-4\fboxrule}
\addtolength\Linewidth{-3\fboxsep}
}
\newenvironment{examplebox}{\par\begingroup
   \setlength{\fboxsep}{5pt}\findlength
   \setbox0=\vbox\bgroup\noindent
   \hsize=0.95\linewidth
   \begin{minipage}{0.95\linewidth}\normalsize}
    {\end{minipage}\egroup
    \textcolor{gray20}{\fboxsep1.5pt\fbox
     {\fboxsep5pt\colorbox{gray05}{\normalcolor\box0}}}
    \endgroup\par\noindent
    \normalcolor\ignorespacesafterend}

\newcommand{\RQ}[2]{%
\refstepcounter{RQCounter} \label{#1}
   \textbf{RQ\arabic{RQCounter}.}~#2
}

\newcommand{\RS}[2]{%
\textbf{Result {\ref{#1}}:~}{{#2}}%
}

\newcommand{\eg}[0]{{\it e.g.,\  }}

{\makeatletter
  \gdef\xxxmark{%
  \expandafter\ifx\csname @mpargs\endcsname\relax 
    \expandafter\ifx\csname @captype\endcsname\relax 
      \marginpar{\textcolor{red}{xxx~}}
    \else
      \textcolor{red}{xxx~}
    \fi
  \else
    \textcolor{red}{xxx~}
  \fi}
\gdef\xxx{\@ifnextchar[\xxx@lab\xxx@nolab}
\long\gdef\xxx@lab[#1]#2{{\bf [\xxxmark \textcolor{red}{#2} ---{\sc #1}]}}
\long\gdef\xxx@nolab#1{{\bf [\xxxmark \textcolor{red}{#1}]}}
}

\newcommand{\hide}[1]{}
\newcommand{\theSystem}{\textsc{Macneto}\xspace}

\newcommand{\sysName}{\theSystem}

\newcommand{\org}{original\xspace}
\newcommand{\obf}{obfuscated\xspace}

\newcommand{\reffig}[1]{Figure \ref{#1}}
\newcommand{\reftable}[1]{Table \ref{#1}}
\newcommand{\tablecap}[1]{\textsc{#1}}

\usepackage[turnon]{notes}

\newtheorem{problem}{Problem}
\DeclareMathOperator*{\argmax}{argmax}

\renewcommand{\textrightarrow}{$\rightarrow$}

\begin{document}

\title[Obfuscation Resilient Search through Executable Classification]{Obfuscation Resilient Search through\\Executable Classification}         

\author{Fang-Hsiang Su}
\affiliation{
    \department{Computer Science}
    \institution{Columbia University, USA}
}
\email{mikefhsu@cs.columbia.edu}

\author{Jonathan Bell}
\affiliation{
    \department{Computer Science}
    \institution{George Mason University, USA}
}
\email{bellj@gmu.edu}

\author{Gail Kaiser}
\affiliation{
    \department{Computer Science}
    \institution{Columbia University, USA}
}
\email{kaiser@cs.columbia.edu}

\author{Baishakhi Ray}
\affiliation{
    \department{Computer Science}
    \institution{Columbia University, USA}
}
\email{rayb@cs.columbia.edu}

\begin{abstract}
Android applications are usually obfuscated before release, making it difficult to analyze them for malware presence or intellectual property violations. 
Obfuscators might hide the true intent of code by renaming variables and/or modifying program structures.
It is challenging to search for executables relevant to an obfuscated application for developers to analyze efficiently.
Prior approaches toward obfuscation resilient search have relied on certain structural parts of apps remaining as landmarks, un-touched by obfuscation.
For instance, some prior approaches have assumed that the structural relationships between identifiers are not broken by obfuscators; others have assumed that control flow graphs maintain their structures. 
Both approaches can be easily defeated by a motivated obfuscator.
We present a new approach, \sysName, to search for programs relevant to obfuscated executables leveraging deep learning and principal components on instructions.
\sysName makes few assumptions about the kinds of modifications that an obfuscator might perform.
We show that it has high search precision for executables obfuscated by a state-of-the-art obfuscator that changes control flow.
Further, we also demonstrate the potential of \sysName to help developers understand executables, where \sysName infers keywords (which are from relevant un-obfuscated programs) for obfuscated executables.
\end{abstract}



\maketitle

\section{Introduction}
\label{sec:intro}

Android apps are typically obfuscated before delivery to decrease the size of distributed binaries and reduce disallowed reuse.
Malware authors may take advantage of the general expectation that Android code is obfuscated to pass off obfuscated malware as benign code: obfuscation will hide the actual purpose of the malicious code, and the fact that there is obfuscation will not be surprising, as it is already a general practice.
Hence, there is great interest in obfuscation resilient search: tools that can automatically find program structures (in a known codebase) likely to be similar to the original version of code that has been obfuscated.

Obfuscation resilient search can be used in various automated analyses, for instance, plagiarism detection~\cite{DBLP:conf/scam/Ragkhitwetsagul16} or detecting precise versions of third party libraries~\cite{DBLP:journals/corr/abs-1802-04466} embedded in applications, allowing auditors to identify the use of vulnerable libraries. 
Similarly, obfuscation resilient search can search among un-obfuscated apps to recover identifiers~\cite{deguard} and/or control flow~\cite{Yadegari:2015:GAA:2867539.2867679} of an obfuscated app. 
Obfuscation resilient search can be useful in human-guided analysis, where an engineer inspects applications to determine security risks.

In general, searching for code likely to be similar to an obfuscated program relies on some training set of pairs of un-obfuscated code and its obfuscated counterpart to build a model.
Once trained, the code search engine can match obfuscated code to its original un-obfuscated code when both obfuscated and un-obfuscated versions are in the corpus.
In the more typical use case when only the obfuscated code is at hand, 
and the un-obfuscated version is unknown to the analyst, the code search may be able to find 
known code highly likely to be similar to the un-obfuscated version.

For example, some code search ``deobfuscation'' tools rely on the structure of an application's control flow graph. 
However, they are susceptible to obfuscators that introduce extra basic blocks and jumps to the application's code and can be slow to use, requiring many pair-wise comparisons to perform their task~\cite{dyclink,discovre}.
Using another approach, DeGuard~\cite{deguard} is a state-of-the-art deobfuscator that builds a probabilistic model for identifiers based on the co-occurrence of names. 
While this technique can be very fast to apply (after the statistical model is trained), it may be defeated by obfuscators that introduce new fields among classes.

We present a novel approach for automated obfuscation resilient search for Android apps, using deep learning: \theSystem, which searches at bytecode level, instead of source code.
\sysName leverages a key observation about obfuscation: an obfuscator's goal is to transform how a program looks as radically as possible, while maintaining the original program semantics.
\sysName works by learning lightweight (partial) executable semantics through Principal Component Analysis (PCA).
These PCA models are a proxy for program behaviors that are stable despite changes to the layout of code, the structure of its control flow graph, or any metadata about the app (features assumed stable by other deobfuscators).
\theSystem's deep PCA model is resilient to multiple obfuscation techniques~\cite{proguard,allatori}, including identifier renaming and control flow modifications.

\sysName uses deep learning to train a classifier on known pairs of un-obfuscated/obfuscated apps offline.
This training process allows \theSystem to be potentially applicable to various obfuscators: supporting a new obfuscator using the same kinds of obfuscations would only require a new data set of pairs of the original un-obfuscated apps and the
corresponding obfuscated apps.
Then, these models are saved for fast, online search where an unknown obfuscated executable is projected to principal components via deep learning, and matched to the most similar executables from the known corpus. 


We evaluated \sysName on $1500+$ real Android apps using an advanced obfuscator: Allatori~\cite{allatori}.
Allatori supports name obfuscation (similar to what ProGuard does~\cite{proguard}) and also control flow obfuscations, e.g., it changes the standard Java constructions for loops, conditional and other branching instructions.
\sysName achieves good search precision, about $80\%$, to retrieve relevant code given unknown obfuscated executables. It significantly outperforms two baseline approaches without deep learning.

The contributions of this paper are:
\begin{itemize}[leftmargin=*]
	\item A new approach to conduct obfuscation resilient code search leveraging deep learning and principal components (features) on bytecode.
	\item A new approach to automate classification of programs with similar semantics.
	\item An evaluation of our tool on an advanced obfuscator.
	\item An open source implementation of \theSystem \cite{macneto}.
\end{itemize}

\section{Background}
\label{sec:motiv}
In general, obfuscators make transformations to code that result in an equivalent execution, despite structural or lexical changes to the code --- generating code that looks different, but behaves similarly.
Depending on the intended purpose (e.g., hiding a company's intellectual property, disguising malware, or minimizing code size), a developer may choose to use a different kind of obfuscator.
These obfuscations might include lexical transformations, control transformations, and data transformations \cite{Collberg:2002:WTO:636196.636198}.
Obfuscators might choose to apply a single sort of transformation, or several.

Lexical transformations are typically employed by ``minimizing'' obfuscators (those that aim to reduce the total size of code\Space{ for distribution}).
Lexical transformations replace identifiers (such as method, class or variable names) with new identifiers.
Since obfuscators are applied only to individual apps, they must leave identifiers exposed via public APIs unchanged.

Control transformations can be significantly more complex, perhaps inlining code from several methods into one, splitting methods into several, reordering statements, adding jumps and other instructions \cite{Low:1998:PJC:332084.332092,6661334}.
Control transformations typically leverage the limitations of static analysis: an obfuscator might add additional code to a method, with a jump to cause the new code to be ignored at runtime.
However, that jump might be based on some complex heap-stored state which is tricky for a static analysis tool to reason about.

Finally, data transformations might involve encoding data in an app or changing the kind of structure that it's stored in.
For instance, an obfuscator might encrypt strings in apps so that they can't be trivially matched, or change data structures. 
For encrypting/decrypting strings, an obfuscator can inject additional helper methods into programs \cite{allatori}.


In this paper we define the obfuscation resilient search problem as follows. 
A developer/security analyst has access to a set of \obf executables, and her job is to identify executables similar with its \org version in the existing codebase. 
The developer/analyst can analyze the \obf executable with its similar ones to identify malware variants, which becomes a significantly easier problem. 
Thus, our task is similar to a search-based deobfuscator DeGuard~\cite{deguard}.


We assume that obfuscators can make lexical, control, and data transformations to code.
We do not base our search model on any lexical features, nor do we base it on the control flow structure of or string/numerical constants in the code.
When inserting additional instructions and methods, we assume that obfuscators have a limited vocabulary of no-op code segments to insert.
We assume that there are some patterns (which need not be pre-defined) that our deep learning approach can detect.
\sysName relies on a training phase that teaches it the rules that the obfuscator follows: if the obfuscator is random with no pattern to the transformations that it makes, then \sysName would be unable to apply its search model to other obfuscated apps.
We imagine that this is a reasonable model: an adversary would have to spend an incredible amount of resources to construct a truly random obfuscator.

\begin{figure*}[ht!pb]
	\centering
	\includegraphics[width=6.8in]{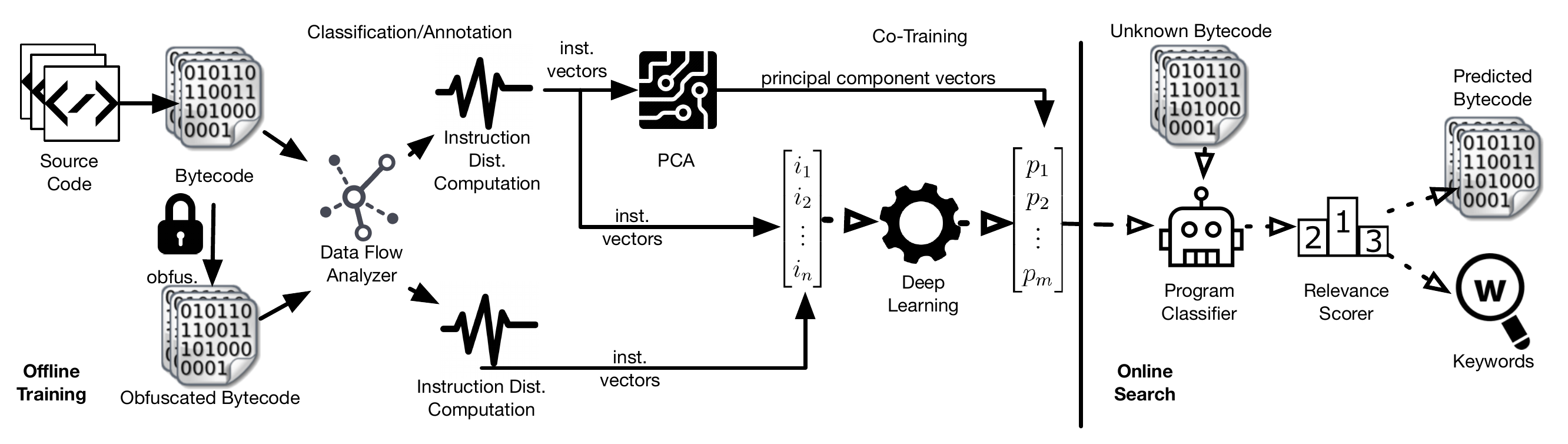}
	\vspace{-18pt}
	\caption{The system architecture of \theSystem, which consists of four stages: instruction distribution, principal component (feature) analysis on bytecode, deep-learning for classifying programs and online scoring, to deobfuscate Android executables.}
	\label{fig:main}
\end{figure*}
\section{\theSystem Overview}
\label{sec:sys}

From a set of \obf APKs, \sysName intends to identify the relevant executables to the original version of a given \obf APK. 
Here we describe an overview of \theSystem. 

Although obfuscators may perform significant structural and/or naming transformations, the semantics of a program before and after obfuscation remain the same.   
\sysName leverages such semantic equivalence between an original program executable and its obfuscated version at the granularity of individual methods.
The semantics of a program executable are the summation of each individual method that it has. 
The semantics of a program executable are captured as the hidden principal components of its machine code instead of human texts such as identifier names in methods and/or descriptions of this program.  
By construction, an obfuscated program/application is semantically equivalent to its original, un-obfuscated version that it is based on. 
\sysName assumes that the principal component vector of an obfuscated application will match those of the original application. 
In its learning phase, \sysName is provided a training set of android applications (APKs), which are labeled pairs of obfuscated and original versions.
Once training is complete, \sysName can be presented with an arbitrary number of obfuscated applications, and for each return suggested applications from its codebase (that it had been trained on) that are similar to the unknown original application.
In the event that the original version happens to exist in the corpus, \sysName will match the obfuscated application with its original version.

\sysName utilizes a four stage approach: 

(i)  {\em Computing Instruction Distribution.} For an application executable (original or obfuscated), \sysName parses each method as a distribution of instructions. 
An application executable can then be represented as the summation of all of its methods, which is also a distribution of instructions.
The instruction distribution of an application executable is analogous to the term frequency vector of a document, where we treat each application executable as a document.

(ii) {\em Principal Component Analysis.} Identifies principal components \cite{pca} from the instruction distribution of the \org app. 
These principal components are used as a proxy for app semantics. 
The same PCA model is used later to annotate the corresponding \obf application\Space{ as well}. 

(iii) {\em Learning.} Uses a three-layered Artificial Neural Network (ANN) \cite{McCulloch1943} where the input is the instruction distribution of an application executable (\org and \obf), and the output layer is the corresponding principal component vector of the \org application. 
\theSystem uses this three-layered ANN as a program classifier that maps an \org application and its \obf version to the same class represented by principal component vector. 
This is the training phase of the ANN model. 
Such model can be pre-trained.

(iv) {\em Obfuscation Resilient Search.} This is the testing phase of the ANN model. 
It operates on a set of \org and \obf applications that form our testing set.
Given an \obf application, the above ANN model tries to infer its principal component vector; \sysName then finds a set of un-obfuscated applications with similar principal component vectors and ranks them as possible deobfuscated candidates. 

Figure \ref{fig:main} shows a high level overview of \theSystem's approach for conducting obfuscation resilient application search. The first three stages occur offline and can be pre-trained. 

\begin{figure}[!htpb]
	\centering
	\includegraphics[scale=0.6]{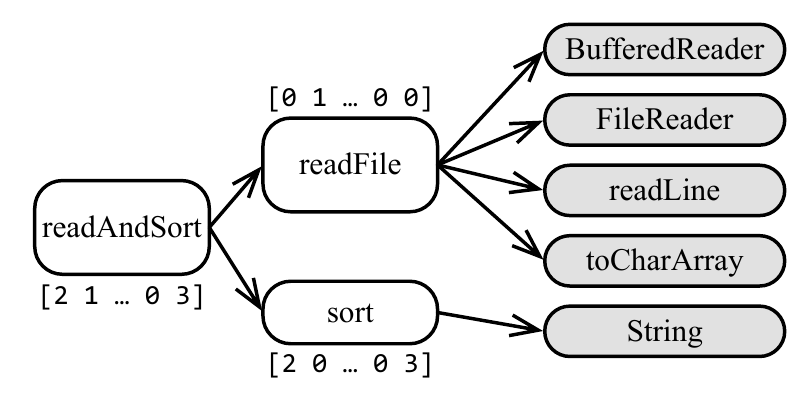}
	\caption{The \texttt{readAndSort} application and two of its methods, \texttt{readFile} and \texttt{sort}.}
	\label{fig:callgraph}
\end{figure}

Consider the example \texttt{readAndSort} program shown in \reffig{fig:callgraph}, assuming that this is an Android app that we are using to train \sysName.
To compute the instruction distribution of \texttt{readAndSort} application, \theSystem first recruits the data flow analysis to identify all possible methods which may be invoked at runtime, which are \texttt{readFile} and \texttt{sort}.
The instruction distributions of these two callee methods will be incorporated into \texttt{readAndSort}.
Then \theSystem moves to the next step, applying Principal Component Analysis (PCA) \cite{pca} on the instruction distributions of all applications including \texttt{readAndSort} in the training app set.
The result of this step is a vector containing the value/membership that a method has toward each principal component: a Principal Component Vector (PCV).

Our insight is that while some instructions in our feature set can be correlated, PCA can help us convert these instructions into orthogonal features.
Furthermore, we can also understand which components are more important to classify application executables. 
These components can help drastically reduce the query time of \sysName to search for similar executables.
\sysName annotates both the original and obfuscated versions of this application with this same PCV.
This annotation process allows our learning phase to predict similar PCVs for an application and its obfuscated version, even their instruction distributions are different.

\section{\theSystem Approach}
\label{sec:mt}

This section describes the four stages of \theSystem in detail, illustrating our several design decisions. 
We have designed \sysName to target any JVM-compatible language (such as Java), and evaluate it on Android apps. 
\sysName works at the level of Java bytecode; in principle, its approach could be applied to other compiled languages as well.
In this paper, executable/binary actually means Java bytecode executable and machine code means Java bytecode.  

\subsection{Computing Instruction Distribution} 
\label{subsubsec:inst}
We use the instruction distribution (ID) of an application to begin to approximate its behavior.
The ID is a vector that represents the frequencies of important instructions. 
We use a bottom-up approach to compute the ID of an application.
Given an application executable $A_{j}$, its ID is the summation of all possible methods invoked at runtime.
For computing the possible invoked methods of an application, we use FlowDroid \cite{Arzt:2014:FPC:2594291.2594299}, a state-of-the-art tool that uses context-, flow-, field-, and object-sensitive android lifecycle-aware control and data-flow analysis~\cite{Arzt:2014:FPC:2594291.2594299}.
For a method $M_{j}^{k}$ that belongs to $A_{j}$, its instruction distribution can be represented as $ID(M_{j}^{k}) = [freq_{I_{1}}, freq_{I_{2}},...freq_{I_{n}}]_{j}^{k}$, where $n$ represents the index of an instruction and $freq_{I_{n}}$ represents the frequency of the $n_{th}$ instruction in the method $M_{j}^{k}$.
The ID of $A_{j}$ can then be computed by $ID(A_{j}) = \sum_{k} ID(M_{j}^{k})$.
The \texttt{readAndSort} program in \reffig{fig:callgraph} can be an example: the ID of \texttt{readAndSort} is the summation of two possibly invoked methods \texttt{readFile} and \texttt{sort}.
This step is similar to building the term frequency vector for a document. 


\begin{table}[!htbp]
\centering
\scriptsize
\vspace{-5pt}
\setlength{\tabcolsep}{0.5pt}
\caption{\tablecap{\theSystem's instruction set.}}
\vspace{-5pt}
\begin{tabular}{ll}
\toprule
\textbf{Opcode} & \textbf{Description}\\
\midrule
\texttt{xaload} & Load a primitive/object x from an  array\\
\texttt{xastore} & Store a primitive/object x to an array\\
\texttt{arraylength} & Retrieve the length of an array. \\
\texttt{xadd} & Add two primitives of type x on the stack.\\
\texttt{xsub} & Subtract two primitives of type x on the stack.\\
\texttt{xmul} & Multiply two primitives of type x on the stack. \\
\texttt{xdiv} & Divide two primitives of type x on the stack. \\
\texttt{xrem} & Compute the remainder of two primitives x on the stack\\
\texttt{xneg} & Negate a primitive of type x on the stack. \\
\texttt{xshift} & Shift a primitive x (type integer/long) on the stack.\\ 
\texttt{xand} & Bitwise-and two primitives of type x (integer/long) on the stack.\\
\texttt{xor} & Bitwise-or two primitives of type x (integer/long) on the stack.\\
\texttt{x\_xor} & Bitwise-xor two primitives x (integer/long) on the stack.\\
\texttt{iinc} & Increment an integer on the stack. \\
\texttt{xcomp} & Compare two primitives of type x on the stack\\
\texttt{ifXXX} & Represent all conditional jumps.\\
\texttt{xswitch} & Jump to a branch based on stack index.\\
\texttt{android\_apis} & The APIs offered by the Android framework\\
\bottomrule
\end{tabular}
\vspace{-10pt}
\label{tab:insts}
\end{table}

\sysName considers various bytecode operations as individual instructions (e.g., adding or subtracting variables), as well as a variety of APIs provided by the Android framework. 
Android API calls provide much higher level functionality than simple bytecode operators, and hence, including them as ``instructions'' in this step allows \sysName to capture both high and low level semantics.
However, including too many different instructions when calculating the distribution could make it difficult to relate near-miss semantic similarity.

To avoid over-specialization, \sysName groups very similar instructions together and represents them with a single word, as shown in Table \ref{tab:insts}. 
For instance, we consider all instructions for adding two values to be equivalent by abstracting away the difference between the instruction \texttt{fadd} for adding floating point numbers and the instruction \texttt{iadd} for adding integers.

To calculate these instruction distributions, \sysName uses the ASM Java bytecode library \cite{ASM}, and Dex2Jar \cite{dex2jar}.
This allows \sysName to search for Android apps (which are distributed as APKs containing Dex bytecode), while only needing to directly support Java bytecode.
For collecting Android APIs, we analyze the core libraries from Android API level $7$ to $25$ \cite{android_build}.
Including these Android APIs, \sysName observes $252$ types of instructions in total.

\subsection{Principal Component Analysis on Executable}
\label{subsec:i2m}

Principal Component Analysis \cite{pca} is a statistical technique that project data containing possibly correlated components into an orthogonal feature space.
While we believe that some instructions may have dependencies, PCA helps us project the IDs of all executables onto an orthogonal space, where each dimension is a principal component.
\sysName uses PCA to convert the ID of each application to a principal component vector (PCV):
\begin{equation}
	PCV(A_{j}) = [Val_{P_{1}}, Val_{P_{2}}...Val_{P_{m}}]_{j}
	\label{eq:app}
\end{equation}
, where $A_{j}$ represents the $j_{th}$ application executable in the application set and $P_{m}$ represents the $m_{th}$ principal component. 
$Val_{P_{m}}$ represents the value that the application executable $A_{j}$ has of the principal component $P_{m}$.

PCA accelerates the search time of \sysName compared with a na\"ive approach using the ID directly to search for similar applications.
More details regarding the performance of \sysName can be found in Section \ref{sec:eval}.
To the best of our knowledge, \theSystem is the first system to identify principal components of programs from machine code.

In \theSystem, we define $32$ principal components ($m = 32$) and have $252$ types of instructions ($n = 252$) as we listed in~\reftable{tab:insts}. 
Using these $32$ principal components, we generate unique principal component vector (PCV).  
Note that, the dimension of each PCV is the same as the principal component number, i.e. $32$, although the number of PCV can be potentially infinite due to different feature values (see Eq.~\ref{eq:app}). 
Thus, an application $A_{j}$ can have a unique $PCV(A_{j})$ that encodes the value of the application belonging to each principal component.  
$PCV(A_{j})$ becomes the semantic representation of both $A_{j}$ and its obfuscated counterpart $A^{ob}_{j}$. 
We annotate each \org and its \obf application with the corresponding PCV and use them to train our ANN based classifier, which will be discussed in Section~\ref{subsec:class}.
To compute PCV, we use the scikit-learn \cite{sklearn} library on machine code.

In the next two steps, \theSystem aims to search for relevant executables to the original version of an obfuscated executable using a ANN based deep learning technique. 
In the training phase, the ANN learns the semantic relationship between a \org and its \obf executable through their unique PCV. 
Next, in the testing (obfuscation resilient search) phase, given an \obf application executable, ANN retrieves a set of candidate applications having similar PCVs with the \obf application. 
\theSystem then scores these candidate applications and outputs a ranked list of relevant un-obfuscated applications with similar PCVs. 

\subsection{Learning Phase}
\label{subsec:class}

In this step, \theSystem uses an ANN based deep learning technique~\cite{SocherGMN13} to project the low-level features (Instruction Distributions) of applications to a relevant vector of principal components (PCV). 
\theSystem treats PCV as a proxy for program semantics, which should be invariant before and after obfuscation. 
Thus, PCV can serve as a signature (i.e., class) of both \org and \obf applications. 
Given a training application set $T$, \theSystem attempts to project each application $A_{j} \in T$ and its obfuscated counterpart $A^{ob}_{j}$ to the same PCV, i.e., $A_{j} \rightarrow PCV(A_{j}) \leftarrow A^{ob}_{j}$.


Similar deep learning technique is widely adopted to classify data. 
However, most of data comes with pre-annotated classes to facilitate learning. 
For example, Socher et al. \cite{SocherGMN13} uses deep learning to classify images to relevant wordings. 
Such work has benchmarked images accompanied with correct descriptions in words to train such classifiers, \theSystem does not have any similar benchmarks.
However, \sysName \emph{does} have available sets of applications, and has access to obfuscators.
Hence, \sysName builds a training set and co-trains a classifier on both obfuscated and original application executables with \sysName knowing the mapping from each training application to its obfuscated counterpart.
The learning phase of \sysName is relevant to the Deep Structured Semantic Models (DSSM) \cite{Huang:2013:LDS:2505515.2505665}, which projects two related corpuses having the same concepts, e.g., queries and their corresponding documents, onto the same feature space.
DSSM can then maximize the similarity based on the relationship between a query and its corresponding document, which can be constructed by clickthrough rate data.
DSSM offers \sysName a future direction, where we can project the origin version of an execution and all of its obfuscated versions onto the same feature space.

\sysName characterizes each application $A_{j}$ and $A^{ob}_{j}$ by the same principal component vector $PCV(A_{j})$, allowing it to automatically tag each application for training program classifiers. 
Given an unknown obfuscated application, \theSystem can first classify it to relevant PCV, which helps quickly search for similar and/or original applications.

To train such projection/mapping, \theSystem tries to minimize the following objective function
\begin{equation}
\begin{split}
	J(\Theta) = \sum_{A_{j} \in T} \norm{PCV(A_{j}) - l(\theta^{(3)} \cdot g(\theta^{(2)} \cdot f(\theta^{(1)} \cdot A_{j})))}^{2} \\
	+  \norm{PCV(A_{j}) - l(\theta^{(3)} \cdot g(\theta^{(2)} \cdot f(\theta^{(1)} \cdot A_{j}^{ob})))}^{2}
\end{split}
\end{equation}
, where $T$ is a training application set, $PCV(A_{j}) \in \mathbb{R}^{n}$ (because \theSystem defines $n$ principal components) and $\Theta = (\theta^{(1)}, \theta^{(2)}, \theta^{(3)})$ defines the weighting numbers for each hidden layer.
For hidden layers, \theSystem uses $relu$ function, where $f(.)$ is the first layer containing $128$ neurons, $g(.)$ is the second layer containing $64$ neurons and $l(.)$ is the third layer containing $32$ neurons. 
\theSystem uses the Adam \cite{adam} solver to solve this objective function.
\theSystem build the ANN on the frameworks of tensorflow \cite{tf, Abadi:2017:CMT:3088525.3088527} and keras \cite{keras}.

\subsection{Obfuscation Resilient Search}
\label{subsec:deob}
Taking an obfuscated application executable as a query, \theSystem attempts to locate which un-obfuscated application(s) in the codebase are mostly similar with it.
The ANN in \theSystem can effectively infer the principal component vector (PCV) of an unknown obfuscated executable and then locate a set of un-obfuscated candidates having similar PCVs measured by the cosine similarity.
Given two application executables, their cosine similarity is defined as
\begin{equation}
	similarity = \frac{PCV(A_{i}) \cdot PCV(A_{j})}{\norm{PCV(A_{i})} * \norm{PCV(A_{j})}}
\label{eq:cosine}
\end{equation}
, where $A_{i}$ and $A_{j}$ are two application executables.

Before we detail the procedure of obfuscation resilient search, we define the terminology that we will use as follows.
\begin{itemize}[leftmargin=*]
	\item $Tr$: The training application executable set.
	\item $Te$: The testing application executable set.
	\item $Tr_{or}$: The training original application set, which is a subset of $Tr$. This is also the search space in our evaluation.
	\item $Tr_{ob}$: The training obfuscated application set, which is a subset of $Tr$ and the obfuscated counterpart of $Tr_{or}$.
	\item $Te_{or}$: The testing original application set, which is a subset of $Te$.
	\item $Te_{ob}$: The training obfuscated application set, which is a subset of $Te$ and the obfuscated counterpart of $Te_{or}$.
	\item $A_{j}$: The $j_{th}$ application executable.
	\item $A_{j}^{ob}$: The obfuscated counterpart of $A_{j}$.
	\item $\sim A_{j}$: A similar application executable of $A_{j}$, where the similarity is measured by their PCV via Eq. \ref{eq:cosine}.
	\item $\{\sim A_{j}\}$: A list of similar application executables sorted by their cosine similarity with $A_{j}$.
\end{itemize}

In our evaluation, we split all Android application executables and their obfuscated counterparts into a training set $Tr$ and a testing set $Te$.
For the training purpose, both $Tr_{or}$ and $Tr_{ob}$ are used to construct the ANN, where only $Tr_{or}$ is recruited for building the PCA model.
For the testing purpose, we use an $A_{j}^{ob}$ in $Te_{ob}$ as a query to search for $n$ ($n = 10$ in this paper) similar application executables $\{\sim A_{j}^{ob}\}$ in $Tr_{or}$.

The original version of $A_{j}^{ob}$, $A_{j}$ in $Te_{or}$ is not in the search space, $Tr_{or}$.
Thus,  to verify the efficacy of \theSystem, we use $A_{j}$ as a query to search for the closest executable $\sim A_{j}$ in $Tr_{or}$, as the groundtruth.
We then check the ranking of $\sim A_{j}$ in $\{\sim A_{j}^{ob}\}$.
By this procedure, we evaluate the search performances of three systems including \theSystem, which will be discussed in Section \ref{sec:eval}.

\section{Evaluation}
\label{sec:eval}

To evaluate the performance of \sysName, we design two large scale experiments to address two research questions based on an advanced obfuscator Allatori \cite{allatori} as follows.

\begin{itemize}[leftmargin=*]
	\item \textbf{RQ1} Executable search: Given an unknown application executable that is obfuscated using lexical, control and data transformation, how accurately can \sysName search for relevant un-obfuscated executables?
	\item \textbf{RQ2} Executable understanding: Given an unknown application executable without source code and  text description, can \sysName infer meaningful keywords for developers/program analyst to understand its semantics?
\end{itemize}

We selected the Allatori obfuscator based on a recent survey of Android obfuscators for its complex control and data-flow modification transformations \cite{Backes:2016:RTL:2976749.2978333}. 
We performed our evaluation on the most recent version at time of experimenting: Allatori 6.5.
To judge \sysName's precision for obfuscation resilient search, we needed a benchmark of plain apps (that is, not obfuscated) from which we could construct training and testing sets.
We used the $1,559$ Android apps from the F-Droid repository as experimental subjects \cite{fdroid}.

We first split these apps into a training set and a testing set and then systematically obfuscate each of them.
Both the original and obfuscated training sets are used to train the program classifier using the first three steps outlined in Section \ref{sec:mt}.
To evaluate the obfuscation resilient search precision of \theSystem, we follow the procedure in Section \ref{subsec:deob} to compare the search results given an obfuscated application $A_{j}^{ob}$ as a query and its original version $A_{j}$ as a query.

As a baseline, we compare \sysName with two approaches as follows:
(1) Na\"ive approach: Calculates the similarity between two applications based on their instruction distributions (IDs) described in Section \ref{subsubsec:inst} without PCA and deep learning.
(2) Pure PCA approach: Calculates the similarity between two applications based on their PCV computed solely by PCA without deep learning.

The major difference between \sysName and the pure PCA approach is that while \sysName uses $Tr_{or}$ to build a PCA model and use it to annotate both $Tr_{or}$ and $Tr_{ob}$ for deep learning, the pure PCA approach use the whole training set ($Tr_{or} + Tr_{ob}$) to build a PCA model without deep learning to transform ID of an application executable to PCV.
The key insight here is that we believe an application $A_{j}$ and its obfuscated counterpart $A_{j}^{ob}$ should share the same semantic classification (PCV), but the pure PCA approach cannot guarantee this invariance. 
The PCVs of $A_{j}$ and $A_{j}^{ob}$ can be different by pure PCA, because their IDs can be different.
This is why \sysName first uses PCA to compute the PCV for $A_{j}$ and use the same PCV to annotate $A_{j}^{ob}$.
The power of deep learning can then help \sysName recognize and search for similar application executables given an unknown obfuscated application.
In our evaluations, we observe that the pure PCA approach can provide good search precision, but \sysName with deep learning can achieve even better precision.

\subsection{Evaluation Metrics}
We use two metrics to evaluate \theSystem's performance to conduct obfuscation resilient search: Top@K and Mean Reciprocal Rank (MRR).
By this procedure, we evaluate the search performances of three systems including \theSystem, which will be discussed in Section \ref{sec:eval}.
\begin{itemize}[leftmargin=*, topsep=3pt]
	\item $Search(A_{j}, Tr_{or}, n)$: Given an application executable $A_{j}$, $Search(.)$ retrieves the most $n$ ($n = 10$ in this paper) similar application executables in the search space, which is the training original application set $Tr_{or}$ in this paper.
	\item $Best(A_{j}, Tr_{or})$: $Best(.)$ is a specialized version of $Search(.)$ to retrieve the most similar application executable ($n = 1$) in the search space.
	\item $Rank(A_{k}, \{A_{l}\}, K)$: Given an application executable $A_{k}$, $Rank(.)$ return $1$ if the ranking of $A_{k}$ in the application list $\{A_{l}\}$ is higher than or equal to $K$.
	If $A_{k}$ is not in $\{A_{l}\}$ or its ranking is lower than $K$, $Rank(.)$ returns $0$.
\end{itemize}

The definition of Top@K can then be
\begin{equation}
	Top@K = \frac{\sum_{j \in Te_{or}}Rank(Best(A_{j}), Search(A_{j}^{ob}, n), K)}{\left\vert Te_{or} \right\vert}
\end{equation}
, where $A_{j}$ is an application executable, $A_{j}^{ob}$ is its obfuscated counterpart.
The search space is $Tr_{or}$ for both $Search(.)$ and $Best(.)$, so we ignore it to simplify the definition.
In our experiments, we use $K = \{1, 5, 10\}$ to evaluate the system performance.

The definition of MRR is
\begin{equation}
	MRR = \frac{1}{\left\vert Te_{or} \right\vert} \sum_{j \in Te_{or}}\frac{1}{Rank(Best(A_{j}), Search(A_{j}^{ob}, n))}
\end{equation}
, where the $Rank(.)$ here returns the ranking of $Best(A_{j})$ in $Search(A_{j}^{ob}, n)$ directly.
If $Best(A_{j})$ is not in $Search(A_{j}^{ob})$, $Rank(.)$ returns 0.

\begin{table*}[h]
\scriptsize
\centering
\setlength{\tabcolsep}{1pt}
\caption{\tablecap{Obfuscation resilient search results of Allatori-obfuscated code.}}
\scalebox{0.98}{
\begin{tabular}{p{0.8cm}p{1.8cm}p{1.3cm}p{1.3cm}p{1.3cm}p{1.3cm} p{1.4cm}p{1.3cm}p{1.3cm}p{1.3cm}p{1.3cm}p{1.3cm}p{1.3cm}}
\toprule
                                   & \textbf{\#Training}        & \textbf{\#Training}   & \textbf{\#Testing} &\textbf{\#Testing} &                          & \textbf{Training}  & \textbf{Query}  &                         &                          &                            &                       &                      \\
\textbf{Exp}                    & \textbf{APKs}               & \textbf{Methods}     & \textbf{APKs}       &\textbf{Methods} & \textbf{System} & \textbf{Time (sec)}       & \textbf{Time (sec)}    & \textbf{Top@1} & \textbf{Top@5} & \textbf{Top@10} & \textbf{MRR} & \textbf{Boost@1}\\
\midrule
\multirow{3}{*}{$\#1$} & \multirow{3}{*}{$1359 + 1359$} & \multirow{3}{*}{$1.14M$} & \multirow{3}{*}{$200 + 200$} & \multirow{3}{*}{$152K$}  &  \theSystem & $2786.72$ & $24.66$ & $0.825$ & $0.96$ & $0.965$ & $0.89$  & +22.22\%\\
                                   &                                                   &                                          &                                       &                                           &   PCA           & $0.0357$   & $20.49$ & $0.78$ & $0.925$ & $0.97$ & $0.85$  & +15.56\%\\
                                   &                                                   &                                          &                                       &                                           &   Na\"ive       & N/A            & $66.55$ & $0.675$ & $0.91$ & $0.94$ & $0.78$  & N/A\\
\midrule
\multirow{3}{*}{$\#2$} & \multirow{3}{*}{$1359 + 1359$} & \multirow{3}{*}{$1.08M$} & \multirow{3}{*}{$200 + 200$} & \multirow{3}{*}{$214K$}  &  \theSystem & $2791.56$ & $24.88$ & $0.82$ & $0.94$ & $0.955$ & $0.87$ & +18.84\%\\
                                   &                                                   &                                         &                                        &                                           &   PCA           & $0.0357$   & $20.73$ & $0.75$ & $0.925$ & $0.95$ & $0.86$ & +8.7\%\\
                                   &                                                   &                                         &                                        &                                           &   Na\"ive       & N/A            & $67.5$ & $0.69$ & $0.91$ & $0.94$ & $0.79$ & N/A\\
\midrule
\multirow{3}{*}{$\#3$} & \multirow{3}{*}{$1359 + 1359$} & \multirow{3}{*}{$1.16M$} & \multirow{3}{*}{$200 + 200$} & \multirow{3}{*}{$137K$}   &  \theSystem & $2826.57$ & $24.76$ & $0.8$ & $0.915$ & $0.955$ & $0.86$ & +22.13\%\\
                                   &                                                   &                                         &                                      &                                            &   PCA           & $0.0352$     & $20.72$ & $0.73$ & $0.94$ & $0.975$ & $0.82$ & +11.45\%\\
                                   &                                                   &                                         &                                      &                                            &   Na\"ive       & N/A            & $66.57$ & $0.655$ & $0.905$ & $0.935$ & $0.77$ & N/A\\
\midrule
\multirow{3}{*}{$\#4$} & \multirow{3}{*}{$1359 + 1359$} & \multirow{3}{*}{$1.15M$} & \multirow{3}{*}{$200 + 200$} & \multirow{3}{*}{$143K$}   &  \theSystem & $2848.41$ & $24.68$ & $0.755$ & $0.905$ & $0.925$ & $0.82$ & +16.15\%\\
                                   &                                                   &                                         &                                      &                                            &   PCA           & $0.0349$     & $20.71$ & $0.715$ & $0.915$ & $0.95$ & $0.80$ & +10\%\\
                                   &                                                   &                                         &                                      &                                            &   Na\"ive       & N/A            & $66.7$ & $0.65$ & $0.865$ & $0.9$ & $0.79$ & N/A\\
\midrule
\multirow{3}{*}{$\#5$} & \multirow{3}{*}{$1359 + 1359$} & \multirow{3}{*}{$1.14M$} & \multirow{3}{*}{$200 + 200$} & \multirow{3}{*}{$157K$}   &  \theSystem & $2842.08$ & $24.72$ & $0.84$ & $0.945$ & $0.96$ & $0.89$ & +15.86\%\\
                                   &                                                   &                                         &                                      &                                            &   PCA           & $0.0336$   & $20.7$ & $0.765$ & $0.93$ & $0.94$ & $0.80$ & +5.5\%\\
                                   &                                                   &                                         &                                      &                                            &   Na\"ive       & N/A            & $67.37$ & $0.725$ & $0.89$ & $0.935$ & $0.79$ & N/A\\
\midrule
\multirow{3}{*}{$\#6$} & \multirow{3}{*}{$1359 + 1359$} & \multirow{3}{*}{$1.14M$} & \multirow{3}{*}{$200 + 200$} & \multirow{3}{*}{$152K$}   &  \theSystem & $2866.28$ & $24.77$ & $0.795$ & $0.95$ & $0.96$ & $0.86$ & +16.05\%\\
                                   &                                                   &                                         &                                      &                                            &   PCA           & $0.0342$   & $20.65$ & $0.72$ & $0.95$ & $0.97$ & $0.85$ & +5.1\%\\
                                   &                                                   &                                         &                                      &                                            &   Na\"ive       & N/A            & $66.64$ & $0.685$ & $0.865$ & $0.935$ & $0.77$ & N/A\\
\midrule
\multirow{3}{*}{$\#7$} & \multirow{3}{*}{$1359 + 1359$} & \multirow{3}{*}{$1.13M$} & \multirow{3}{*}{$200 + 200$} & \multirow{3}{*}{$166K$}   &  \theSystem & $2866.14$ & $24.81$ & $0.82$ & $0.915$ & $0.935$ & $0.86$ & +29.13\%\\
                                   &                                                   &                                         &                                      &                                            &   PCA           & $0.0347$   & $20.65$ & $0.735$ & $0.94$ & $0.95$ & $0.82$ & +15.75\%\\
                                   &                                                   &                                         &                                      &                                            &   Na\"ive       & N/A            & $66.68$ & $0.635$ & $0.87$ & $0.915$ & $0.74$ & N/A\\
\midrule
\multirow{3}{*}{$\#8$} & \multirow{3}{*}{$1400 + 1400$} & \multirow{3}{*}{$1.13M$} & \multirow{3}{*}{$159 + 159$} & \multirow{3}{*}{$167K$}   &  \theSystem & $2958.32$ & $20.192$ & $0.79$ & $0.90$ & $0.93$ & $0.84$ & +4.2\%\\
                                   &                                                   &                                         &                                      &                                            &   PCA           & $0.039$   & $16.94$ & $0.73$ & $0.91$ & $0.93$ & $0.81$ & -3.3\%\\
                                   &                                                   &                                         &                                      &                                            &   Na\"ive       & N/A            & $54.5$ & $0.75$ & $0.90$ & $0.93$ & $0.82$ & N/A\\
\bottomrule
\end{tabular}
}
\label{tab:all_allatori}
\\ \flushleft
{\underline{Column Description:} Exp: Experiment ID; Train APKs and Test APKs: numbers of training and testing APKs, respectively; Training Methods and Testing Methods: denote the method numbers belonging to Training APKs and Testing APKs, respectively; System: system under evaluation; Training Time: the time that each system spends to training the model for each experiment; Query Time: the time that each system spends to search for relevant executables given an unknown and obfuscated executable;  Top@K: the percentage of queries (executables), where their original versions are retrieved and ranked by each system at or better than $K_{th}$ position; MRR: Mean Reciprocal Ranking of each system;  Boost@1:  the enhancements achieve by \theSystem and the pure PCA approach over the na\"ive approach on precision (Top@1).}
\end{table*}

\subsection{Executable Search}
\label{subsec:ee_search}
In this section, we answer the research question:
\RQ{rq1}{Given an unknown application executable that is obfuscated using lexical, control and data transformation, how accurately can \sysName search for relevant un-obfuscated executables?}

Compared with lexical obfuscators like ProGuard \cite{proguard} that mainly focuses on renaming identifiers in programs, Allatori changes control flow and encrypts/decrypts strings via inserting additional methods into programs.
To demonstrate the performance of \theSystem to search for relevant application executables against such advanced obfuscations, we conduct a K-fold analysis ($K=8$) on the Android application set we have.
We first split the $1559$ Android application executables we have into $8$ folds.
For each experiment, we use $7$ folds to train an obfuscation resilient search model and use the other as the testing set (queries).
In total, we trained $8$ models for $8$ experiments, where each application executable will be in the testing set for once and in the training set for $7$ times.
This K-fold analysis can help us verify the robustness of \theSystem, because each executable will be tested.

The overall results of the $8$ models trained by \theSystem and two baseline approaches can be found in \reftable{tab:all_allatori}.
In Table \ref{tab:all_allatori}, the ``Exp'' column represents the experiment ID, where we have $8$ folds (experiments) in total.
The ``Training APKs' and ``Testing APKs'' columns represent the numbers of training and testing apps, respectively.
Note that the size of training apks does not matter to the na\"ive approach, because it simply relies on instruction distributions to search for programs.
The reason that we offer two numbers, e.g., $1359 + 1359$, for the training apks is that the training phase includes both original apks in $Tr_{or}$ and their obfuscated counterparts in $Tr_{ob}$.
The ``Training Methods'' and ``Testing Methods'' columns show the number of methods including in the training apks and testing apks, respectively.
Note that while \theSystem includes all possible invoked methods of executables (apks) to compute instruction distributions, our training and testing works at executable level.
The ``System'' column shows which system we evaluate, where ``PCA'' and ``Na\"ive'' are the two baseline approaches we discussed in Section \ref{sec:eval}.
The ``Training Time'' and ``Query Time" column shows the time consumed by each system to complete the training and testing (query) in seconds. 
The ``Top@K'' columns, where $K=\{1, 5, 10\}$, and the ``MRR'' column is self-explanatory.
The ``Boost@1'' column shows the improvement of \theSystem and the pure PCA approach against the na\"ive approach.

We observe three key findings of \theSystem in \reftable{tab:all_allatori}:
\begin{enumerate}[leftmargin=*, topsep=3pt]
	\item Good effectiveness of obfuscation resilient research: \theSystem can achieve $80+\%$ Top@1 for most experiments.
	\item Effectiveness of the executable classifier trained by deep learning: Compared with the na\"ive and the pure PCA approach without deep learning, \theSystem achieves $17.76\%$ and $8.72\%$ enhancements of Top@1, respectively.
	In our experiment, PCA is an effective technique to extract application semantics, because the pure PCA approach already offers $8.31\%$ enhancement of Top@1 over the na\"ive approach.
	Deep learning helps \theSystem understand executables further ($17.76\%$ enhancement) with PCA.
	\item Query performance by PCA: Compared with the na\"ive approach which searches for similar applications based on $252$ types of instructions, \theSystem and the pure PCA approach searches only by $32$ principal components.
	This leads to great runtime performance of both systems to search for similar application executables: while the na\"ive approach needs $65.09$ seconds in average to process $200$ queries, \theSystem and the pure PCA approach only need $24.09$ and $20.13$ seconds in average.
\end{enumerate}

\RS{rq1}{\theSystem can achieve up to 84\% precision (Top@1) for searching for similar application executables given unknown and obfuscated executables as queries. It significantly outperforms two baseline approaches in precision and MRR.}

\subsection{Executable Understanding}
\label{subsec:eu}
In addition to searching for similar executables, we are interested in exploring the potential of \theSystem to support developers quickly understanding an unknown application executable without human descriptions.
We will answer the research question:
\RQ{rq2}{Given an unknown application executable without source code and  text description, can \sysName infer meaningful keywords for developers/program analyst to understand its semantics?}

We crawled the F-Droid repository \cite{fdroid} to extract the description for each un-obfuscated APK. 
Then we follow the search procedure in Section \ref{subsec:deob} to retrieve $10$ most similar application executables in $Tr_{or}$, given an obfuscated executable $A_{j}^{ob}$.
Among these $10$ similar application executables, we use a TF-IDF \cite{Rajaraman:2011:MMD:2124405} model to extract a set of keywords from their descriptions.
From this set of keywords, we select top $10$ keywords having the highest TF-IDF values.
These selected keywords become the human semantic that we predict for an application executable $A_{j}^{ob}$ without human descriptions.
To verify the correctness of these keywords, we compare with the real description of $A_{j}$, the original version of $A_{j}^{ob}$.

To conduct this experiment, we randomly select $1,539$ application executables as the training set and use the rest $20$ as the testing set.
We then manually compare the predicted keyword set of $A_{j}^{ob}$ with the real description of its original version $A_{j}$.
Due to the page limitation, we are not able to offer all the results.
Instead, we list some interesting cases that \theSystem precisely infers meaningful words to describe an unknown and obfuscated executable without any descriptions.
\begin{itemize}[leftmargin=*, topsep=3pt]
	\item \texttt{com.platypus.SAnd}: The partial description of this executable is ``Use your phones sensors...to show your current \textbf{orientation}, height and air pressure...''. Given the obfuscated version of this executable as a query, \theSystem infers a relevant keyword \textbf{coordinates}, where the pure PCA and the na\"ive approaches fail to offer meaningful keyword.
	\item \texttt{se.danielj.geometridestroyer}: This application executable is a game app \cite{geometri-destroyer}, but its description does not mention any words relevant to ``game'': ``Remove the green objects but don't let the blue objects touch the ground''. 
	However, \theSystem predicts two relevant words of ``game'', \textbf{game} and \textbf{libgdx} (which is a framework to develop Android game app) to describe this executable.
	The na\"ive approach predicts these two keywords as well, where the pure PCA approach does not offer any relevant words.
	\item \texttt{net.bierbaumer.otp\_authenticator}: This application executable offers a two-factor authentication functionality to users, which users can scan QR code to log in. 
	The description of this application is ``OTP Authenticator is a two-factor \textbf{authentication}...Simply scan the \textbf{QR} code...''.
	While the na\"ive approach predicts a relevant word \textbf{privacy}, \theSystem infers two relevant words, \textbf{QR} and \textbf{security}, which precisely describe this app.
	The pure PCA approach fails to offer any relevant words.
\end{itemize}

While \theSystem is able to provide at least one meaningful keyword for $14/20$ obfuscated executables in the testing set, the na\"ive approach and the pure PCA approach can only achieve $7/20$ and $4/20$, respectively.
Determining the relevance of a keyword to an executable can be subjective, so we plan to conduct user studies to examining the efficacy of \theSystem on the executable understanding in the future.

\RS{rq2}{\theSystem has the potential to infer meaningful human words for developers/program analysts to understand an unknown executable, even it is obfuscated and has no human descriptions.}

\subsection{Discussions}
We discuss the limitations of \sysName and the potential solutions as follows.
In this paper, we have shown that \sysName can search for relevant executables, even they are obfuscated by control flow transformation and anonymization.
However, while \sysName relies on instruction distributions (ICs) to search for relevant executables, adding noisy instructions that change the IC dramatically without affecting the functionality of an executable may not be handled by \sysName.
If two executables have totally different functionalities, but their ICs become similar after adding some noises, \sysName will falsely detect them as similar programs.
A potential solution is to leverage data flow analysis at instruction level to collect useful instructions that can influence outputs of executables to compute ICs for \sysName.
Another relevant concern is that while we believe that \sysName can be a generic approach to tackle various obfuscators, we only discuss one in this paper.
We plan to collect more obfuscators to conduct further experiments to prove the efficacy of \sysName in general.

While optimizing the hyper parameters, such as the layer number, of deep learning technique is out of the scope of this paper, we want to discuss the trade-off between precision and training time of \sysName.
Adding more layers in \sysName can possibly enhance the search precision, but this will also increase training time.
For deciding the layer number, we follow \cite{SocherGMN13} to start from fewer layers ($3$ in this paper), which facilitate us verifying the effectiveness of \sysName in a timely fashion.
In the future, after we collect more obfuscators, we believe that \sysName would definitely need more layers to classify executables.

In this paper, we use principal components as a proxy (representation) of executable behaviors.
We plan to explore more techniques, such as autoencoders \cite{Baldi:2011:AUL:3045796.3045801}, to extract and represent executable behaviors.

\section{Related Work}
\label{sec:related}


Although in a programming language identifier names can be arbitrary,  real developers usually use meaningful names for program comprehension~\cite{liblit2006cognitive}.
Allamanis et al.\cite{allamanis2014learning} reported that code reviewers often suggest to modify identifier names before accepting a code patch. 
Thus, in recent years, naming convention of program identifiers drew significant attention for improving program understanding and maintenance~\cite{allamanis2014learning, butler2015investigating, lawrie2006s, takang1996effects, arnaoudova2016linguistic, martin2009clean}.
Among the identifiers, a good method name is particularly helpful because they often summarize the underlying functionalities of the methods~\cite{host2009debugging,allamanis2015suggesting}.
Using a rule-based technique, Host et al.~\cite{host2009debugging} inferred method names for Java programs using the methods' arguments, control-flow, and return types.
In contrast, Allamanis et al. used a neural network model for predicting method names of Java code~\cite{allamanis2015suggesting}.
Although these two studies can suggest better method names in case of naming bugs, they do not look at the obfuscated application executables that can even change the structure of the program.

JSNice \cite{Raychev:2015:PPP:2676726.2677009} and DeGuard \cite{deguard} apply statistical models to suggest names and identifiers in JavaScript and Java code, respectively.
These statistical models work well against so called ``minimizers'' --- obfuscators that replace identifier names with shorter names, without making any other transformations.
These approaches may not be applied to obfuscators that modify program structure or control flow.

While \sysName uses PCA as a proxy for application behavior, a variety of other systems use input/output behavior \cite{Jiang:2009:AMF:1572272.1572283,hitoshiio,Juergens:2010:CSB:1955601.1955971}, call graph similarity \cite{dyclink,discovre}, or dynamic symbolic execution \cite{Meng:2016:SMA:2931037.2931043,Li:2016:MCB:2889160.2889204,6671332} at method level.
\sysName is most similar to systems that rely on software birthmarks, which use some representative components of a program's execution (often calls to certain APIs) to create an obfuscation-resilient fingerprint to identify theft and reuse \cite{Tamada04designand,Schuler:2007:DBJ:1321631.1321672,McMillan:2012:DSS:2337223.2337267,appcontext,mudflow,Linares-Vasquez:2014:UML:2617668.2617703}.
One concern in birthmarking is determining which APIs should be used to create the birthmark: perhaps some API calls are more identifying than others.
\theSystem extends the notion of software birthmarking by using deep learning to identify patterns of APIs and instruction mix, allowing it be an effective executable search engine.
\section{Conclusion}
\label{sec:conclusion}
We present \theSystem, which leverages deep learning and PCA techniques at the executables level (bytecode) to search for programs (in a known corpus) similar to a given obfuscated executable.
In a large scale experiment, we show that \theSystem can achieve up to $84\%$ precision to search for relevant Android executables even when the query executable is obfuscated by anonymization and control flow transformation.
Compared with a na\"ive approach relying on instruction distribution to search for relevant executables, \theSystem improves search precision by up to $29\%$.
We also show the potential of \theSystem to infer meaningful keywords for unknown executables without human descriptions.

\section*{Acknowledgements}
\label{sec:ack}
We would like to thank all of our reviewers and our shepherd, Alvin Cheung, for their valuable comments and suggestions.
This work was done while Fang-Hsiang Su was a PhD candidate at Columbia University.
The Programming Systems Laboratory is funded in part by NSF CNS-1563555.

\bibliography{mt,dyclink,missing}

\end{document}